\documentstyle[11pt]{article}

\newlength{\minitwocolumn}
\catcode`\@=11
\def\relaxnext@{\let\next\relax}
\font\tenmsy=msym10 scaled\magstep1
\font\sevenmsy=msym7 scaled\magstep1
\font\fivemsy=msym5  scaled\magstep1
\newfam\msyfam
\textfont\msyfam=\tenmsy
\scriptfont\msyfam=\sevenmsy
\scriptscriptfont\msyfam=\fivemsy
\font\teneuf=eufm10 scaled\magstep1
\font\seveneuf=eufm7 scaled\magstep1
\font\fiveeuf=eufm5 scaled\magstep1
\newfam\euffam
\textfont\euffam=\teneuf
\scriptfont\euffam=\seveneuf
\scriptscriptfont\euffam=\fiveeuf
\def\frak{\relaxnext@\ifmmode\let\next\frak@\else
 \def\next{\Err@{Use \string\frak\space only in math mode}}\fi\next}
\def\goth{\relaxnext@\ifmmode\let\next\frak@\else
 \def\next{\Err@{Use \string\goth\space only in math mode}}\fi\next}
\def\frak@#1{{\frak@@{#1}}}
\def\frak@@#1{\noaccents@\fam\euffam#1}
\def\Bbb{\relaxnext@\ifmmode\let\next\Bbb@\else
 \def\next{\Err@{Use \string\Bbb\space only in math mode}}\fi\next}
\def\Bbb@#1{{\Bbb@@{#1}}}
\def\Bbb@@#1{\noaccents@\fam\msyfam#1}
\def\accentfam@{7}
\def\noaccents@{\def\accentfam@{0}}
\catcode`\@=\active

\def\eq#1\endeq{\begin{eqnarray}#1\end{eqnarray}}
\def\eqn#1\endeqn{\begin{eqnarray*}#1\end{eqnarray*}}

\def\Hom{{\rm \Hom}}

\makeatletter
\@addtoreset{equation}{section}

\makeatother

\newtheorem{thm}{Theorem}[section]
\newtheorem{prop}[thm]{Proposition}
\newtheorem{lem}[thm]{Lemma}
\newtheorem{cor}[thm]{Corollary}

\begin{document}
\begin{flushright}
\end{flushright}
\vspace{24pt}
\begin{center}
\begin{Large}
{\bf Maxwell-Bloch equation and\\
Correlation functions for penetrable Bose gas
}
\end{Large}

\vspace{36pt}
T. Kojima\raisebox{2mm}{{\scriptsize 1}$\star$}
and
V. Korepin\raisebox{2mm}{{\scriptsize 2}}

\vspace{6pt}

~\raisebox{2mm}{{\scriptsize 1}}
{\it Research Institute for Mathematical Sciences,
     Kyoto University, Kyoto 606, Japan}

~\raisebox{2mm}{{\scriptsize 2}}
{\it Institute for Theoretical Physics, State University of
New York at Stony Brook,
Stony Brook, NY 11794-3840, U. S. A.}

\vspace{72pt}

\vskip 3mm
\begin{footnotesize}
\noindent 
\raisebox{2mm}{$1$}
kojima@kurims.kyoto-u.ac.jp~~
\raisebox{2mm}{$2$}
korepin@insti.physics.sunysb.edu\\
\noindent\raisebox{2mm}{$\star$}
Research Fellow of the Japan Society
for the Promotion of Science.
\end{footnotesize}
\newpage

\underline{Abstract}

\end{center}

We consider the quantum nonlinear Schr\"odinger equation
in one space and one time dimension.
We are interested in the non-free-fermionic case.
We consider static temperature-dependent correlation functions.
The determinant representation for correlation functions
simplifies in the small mass limit of the Bose particle.
In this limit we describe the correlation functions by the vacuum
expectation value of a boson-valued solution for
Maxwell-Bloch differential equation.
We evaluate long-distance
asymptotics of correlation functions in the small mass limit.
\vspace{24pt}

\newpage

\section{Introduction}
In this paper we consider correlation functions
of exactly solvable models.
Our approach is based on the determinant representation 
of quantum correlation functions \cite{K.B.I.}.
We consider the thermodynamics 
of Bose gas with delta-interaction at finite temperature $T>0$.
The one dimensional Bose gas with delta-function interaction
is described by the canonical Bose fields
$\psi(x)$ and $\psi^{+}(x)$ with the commutation relations :
\begin{eqnarray}
[\psi(x), \psi^{+}(y)]=\delta(x-y),~~
[\psi(x), \psi(y)]=[\psi^{+}(x), \psi^{+}(y)]=0.
\end{eqnarray}
The Hamiltonian of the model is
\begin{eqnarray}
{\it H}=\int dx
\left(\frac{1}{2m}\frac{\partial}{\partial x}\psi^{+}(x) 
\frac{\partial}{\partial x} \psi(x)+
g\psi^{+}(x)\psi^{+}(x)\psi(x)\psi(x)
-h \psi^{+}(x)\psi(x)\right),
\end{eqnarray}
where $m>0$ is the mass, $g>0$ is the coupling constant and $h>0$
is the chemical potential.
The Hamiltonian $H$ acts on the Fock space
with the vacuum vector $\vert vac \rangle$.
The vacuum vector $\vert vac \rangle$ is characterized by
the relation :
\begin{eqnarray}
\psi(x) \vert vac \rangle =0.
\end{eqnarray}
The dual vacuum vector $\langle vac \vert$ is characterized by the relations :
\begin{eqnarray}
\langle vac \vert \psi^{+}(x)=0,~~ \langle vac \mid vac \rangle=1.
\end{eqnarray}
The corresponding equation of motion,
\begin{eqnarray}
i\frac{\partial}{\partial t} \psi=
[\psi, H]=
-\frac{1}{2m}\frac{\partial^2}{{\partial x}^2} \psi 
+2g \psi^{+}\psi \psi
-h \psi,
\end{eqnarray}
is called the quantum nonlinear Schr\"odinger equation
in one space and one time dimension.
The quantum field theory problem is reduced to a quantum
mechanics problem.
It is well known that in the $N$ particle sector the eigenvalue
problem $H \vert \psi_N \rangle = E_N \vert \psi_N \rangle $,
is equivalent to the one described by the quantum mechanics $N$
body Hamiltonian :
\begin{eqnarray}
H_N=-\frac{1}{2m}\sum_{j=1}^N
\frac{\partial ^2}{\partial z_j^2}
+2g \sum_{1 \leq j < k \leq N}
\delta(z_k-z_j)-Nh.
\end{eqnarray}
E.H. Lieb and W. Linger \cite{L.L.} solved the eigenvalue problem
$H_N \psi_N=E_N \psi_N$.
They constructed the eigen-functions
$\psi_N=\psi_N(z_1,\cdots,z_N\vert
\lambda_1,\cdots,\lambda_N)$ by means of
the Bethe Ansatz.
The eigen-function $\psi_N=\psi_N(z_1,\cdots,z_N\vert
\lambda_1,\cdots,\lambda_N)$
depends on the spectral parameter
$\lambda_1<\cdots<\lambda_N$.
The spectral parameters $\lambda_1 < \cdots <\lambda_N$
are determined by the periodic boundary conditions :
\begin{eqnarray}
\psi_N(z_1,\cdots,z_j+L,\cdots,z_N \vert
\lambda_1,\cdots,\lambda_N)
=\psi_N(z_1,\cdots,z_j,\cdots,z_N \vert
\lambda_1,\cdots,\lambda_N),
\end{eqnarray}
which amounts to the Bethe Ansatz equations :
\begin{eqnarray}
e^{i\lambda_j L}
=-\prod_{k=1}^N
\frac{\lambda_j-\lambda_k+2i mg}
{\lambda_j-\lambda_k-2i mg},~~j=1,\cdots,N.
\end{eqnarray}
Here $L>0$ is the size of the box.
The eigenvalue of the Hamiltonian $H_N$ is given by
\begin{eqnarray}
E_N=\sum_{j=1}^N\left(
\frac{1}{2m}\lambda_j^2-h\right).
\end{eqnarray}
E.H. Lieb and W. Linger \cite{L.L.}, \cite{L}
discussed the zero temperature thermodynamic limit.
The ground state and its 
excitations are described by linear integral equations.
C.N. Yang and C.P. Yang \cite{Y.Y.}
discussed the finite temperature thermodynamic limit.
The state of thermodynamic equilibrium is described by
non-linear integral equations.
The density of particles $\rho_p(\lambda)$ and 
the density of holes $\rho_h(\lambda)$ are described by the
following non-linear integral equations :
\begin{eqnarray}
2\pi \rho_t(\lambda)&=&1+\int_{-\infty}^{\infty}
K(\lambda,\mu)\rho_p(\mu)d\mu,\\
D&=&\frac{N}{L}=\int_{-\infty}^{\infty}\rho_p(\mu)d\mu,\\
\varepsilon(\lambda)&=&\frac{\lambda^2}{2m}
-h-\frac{T}{2\pi}\int_{-\infty}^{\infty}K(\lambda,\mu)\ln
\left(1+e^{-\frac{\varepsilon(\mu)}{T}}\right)d \mu,
\end{eqnarray}
where $T>0$ is temperature and $D=\frac{N}{L}$
is the density of particles.
Here the functions $\varepsilon(\lambda)$ and $\rho_t(\lambda)$
are defined by
\begin{eqnarray}
\frac{\rho_h(\lambda)}{\rho_p(\lambda)}=
e^{\frac{\varepsilon(\lambda)}{T}},~~\rho_t(\lambda)
=\rho_p(\lambda)+\rho_h(\lambda).
\end{eqnarray}
The integral kernel $K(\lambda,\mu)$ is defined by
\begin{eqnarray}
K(\lambda,\mu)=\frac{4mg}{(\lambda-\mu)^2+(2mg)^2}.
\label{def:kernel-K}
\end{eqnarray}
Consider the local density operator
$j(x)=\psi^{+}(x)\psi(x)$.
In this paper we consider the mean value of the operator :
\begin{eqnarray}
\exp\left(\alpha Q(x)\right).
\end{eqnarray}
Here $\alpha$ is an arbitrary complex parameter
and $Q(x)$ is the operator of the number of particles on the interval
$[0,x]$ :
\begin{eqnarray}
Q(x)=\int_0^x\psi^{+}(y)\psi(y)dy.
\end{eqnarray}
We are interested in the generating function of
temperature-dependent correlation function
defined by
\begin{eqnarray}
\langle \exp \left(\alpha Q(x)\right) \rangle_T
=\frac{{\rm tr}\left(\exp\left(-\frac{H}{T}\right)
\exp\left(\alpha Q(x)\right)\right)}
{{\rm tr}\left(\exp\left(-\frac{H}{T}\right)\right)}.
\end{eqnarray}
The expectation value
$\langle \exp \left(\alpha Q(x)\right) \rangle_T$
is a remarkable quantity, because
a lot of interesting correlation function can be extracted from
$\langle \exp \left(\alpha Q(x)\right) \rangle_T$.
For example the density correlation function,
\begin{eqnarray}
\langle j(x)j(0) \rangle_T
=\frac{{\rm tr}\left(\exp\left(-\frac{H}{T}\right)j(x)j(0)\right)}
{{\rm tr}\left(\exp\left(-\frac{H}{T}\right)\right)}
\end{eqnarray}
can be derived by
\begin{eqnarray}
\langle j(x)j(0) \rangle_T
=\frac{1}{2}
\frac{\partial^2}{\partial x^2}
\langle Q(x)^2 \rangle_T
=\frac{1}{2}\frac{\partial^2}{\partial x^2}
\left.\frac{\partial^2}{\partial \alpha^2}
\langle \exp \left(\alpha Q(x)\right)\rangle_T
\right|_{\alpha=0}.
\end{eqnarray}
In this paper we are interested in the small mass limit of
the Bose particle :
\begin{eqnarray}
m \to 0,~g \to \infty~~\mbox{such that the product}
~~c=2mg~~ \mbox{is fixed.}
\end{eqnarray}
We want to emphasize that the small mass limit is not a free-fermionic
limit.
The scattering matrix of the particles $\lambda_p$ 
and $\lambda_h$ is equal to
\begin{eqnarray}
S(\lambda_p,\lambda_h)
=\exp \left(-i \delta(\lambda_p,\lambda_h)\right),~~
\lambda_p > \lambda_h,
\end{eqnarray}
where the scattering phase $\delta$ satisfying the following
integral equation :
\begin{eqnarray}
\delta(\lambda_p,\lambda_h)-\frac{1}{2\pi}
\int_{-\infty}^{\infty}K(\lambda_p,\mu)\vartheta(\mu)
\delta(\mu,\lambda_h)
=i \ln\left(\frac{i c+\lambda_p-\lambda_h}
{i c-\lambda_p+\lambda_h}\right).
\end{eqnarray}
Here we used
\begin{eqnarray}
\vartheta(\lambda)=
\frac{1}{1+e^{\frac{\varepsilon(\lambda)}{T}}}
=\frac{\rho_p(\lambda)}{\rho_t(\lambda)}.
\end{eqnarray}
Therefore the small mass limit
is not a free-fermionic limit. 
In the small mass limit we will show that
the expectation value $\langle \exp \left(\alpha Q(x)\right)\rangle_T$
is described by the vacuum expectation value of a boson-valued solution of
the Maxwell-Bloch equation \cite{B.G.Z.}.
The plan of this paper is as follows.
In section 2 we summarize known results of determinant representations
for correlation functions.
In section 3 we consider the small mass limit of 
temperature correlation functions.
The determinant representation for correlation functions simplifies
in the small mass limit.
In section 4 we show that correlation functions can be described by the
vacuum expectation value of
a boson-valued solution of Maxwell-Bloch equation,
in the small mass limit.
In section 5 we evaluate asymptotics of the correlation functions
in the small mass limit.

\begin{section}{Determinant representation with dual fields}

The purpose of this section is to summarize the known results
of determinants representation for temperature correlation functions
\cite{K.B.I.}.
First, we introduce the dual fields
$\phi_j(\lambda), ~(j=1,\cdots,4)$ defined by
\begin{eqnarray}
\phi_j(\lambda)=p_j(\lambda)+q_j(\lambda), ~~(j=1,\cdots,4).
\end{eqnarray}
Here the fields $p_j(\lambda)$ and $q_j(\lambda)$
are defined by the commutation relations :
\begin{eqnarray}
[p_j(\lambda), p_k(\mu)]=[q_j(\lambda), q_k(\mu)]=0,~~
[p_j(\lambda), q_k(\mu)]
=H_{j,k}(\lambda,\mu),~~(j,k=1,\cdots,4).
\end{eqnarray}
Here we used 
\begin{eqnarray}
H_{j,k}(\lambda,\mu)
=
\left(
\begin{array}{cccc}
-1 & 0 & 0 & -1\\
0  & -1& 1 &0 \\
1  &  0&-1 &1 \\
0  & -1& 1 &-1
\end{array}
\right)_{j,k}
\ln \left(h(\lambda,\mu)\right)+
\left(
\begin{array}{cccc}
-1 & 0 & 1 & 0\\
0  & -1& 0 &-1 \\
0  &  1&-1 &1 \\
-1  & 0& 1 &-1
\end{array}
\right)_{j,k}
\ln \left(h(\mu,\lambda)\right),
\end{eqnarray}
where
\begin{eqnarray}
h(\lambda,\mu)=\frac{1}{i c}(\lambda-\mu+i c).
\end{eqnarray}
The dual fields $\phi_j(\lambda)$ commute.
\begin{eqnarray}
[\phi_j(\lambda), \phi_k(\mu)]=0,~~(j,k=1,\cdots,4).
\end{eqnarray}
We introduce the auxiliary Fock space with the auxiliary vacuum vector
$\vert 0 \rangle$.
The auxiliary vacuum vector $\vert 0 \rangle$
is characterized by
\begin{eqnarray}
p_j(\lambda)\vert 0 \rangle=0,~~(j=1,\cdots,4).
\end{eqnarray}
The auxiliary dual vacuum
$\langle 0 \vert$
is characterized by
\begin{eqnarray}
\langle 0 \vert q_j(\lambda)=0,~~(j=1,\cdots,4),
~~\langle 0 \mid 0 \rangle=1.
\end{eqnarray}
We want to emphasize that the dual fields
$\phi_j(\lambda)~(j=1,\cdots,4)$ and the auxiliary Fock space
can be written in terms of the four standard Bose fields
$\psi_j(\lambda),~\psi_j^{+}(\mu),~(j=1,\cdots,4)$
and the standard Fock vacuum $\vert 0 \rangle$
and the dual Fock vacuum $\langle 0 \vert $ :
\begin{eqnarray}
[\psi_j(\lambda), \psi_k^{+}(\mu)]=
\delta_{j,k} \delta(\lambda-\mu),~~
[\psi_j (\lambda), \psi_k (\mu)] =
[\psi_j^{+}(\lambda), \psi_k^{+}(\mu)]=0,~~(j,k=1,\cdots,4).
\end{eqnarray}
\begin{eqnarray}
\psi_j(\lambda)\vert 0 \rangle=0,&~&
\langle 0 \vert \psi_j^{+}(\lambda)=0.
\end{eqnarray}
Actually, the dual fields can be realized by
\begin{eqnarray}
p_j(\lambda)=\psi_j(\lambda),~~
q_k(\mu)=\sum_{l=1}^4\int_{-\infty}^{\infty}
H_{l,k}(\nu,\mu)\psi^{+}_l(\nu)d \nu,~~(j,k=1,\cdots,4).
\end{eqnarray}
Next we prepare two integral operators
$\hat{V}_T$ and $\hat{K}_T$.
The integral operator $\hat{V}_T$ is defined by
\begin{eqnarray}
\left(\hat{V}_Tf\right)(\lambda)=\int_{-\infty}^{\infty}
V_T(\lambda, \mu)f(\mu) d\mu.
\end{eqnarray}
The integral kernel $V_T(\lambda, \mu)$ is defined by
product $V_T(\lambda,\mu)=V(\lambda,\mu)\vartheta(\mu)$.
The first factor $V(\lambda,\mu)$ is defined by
\begin{eqnarray}
V(\lambda,\mu)&=&\frac{1}{c}\left\{
t(\lambda,\mu)+t(\mu,\lambda)\exp\left(-i x(\lambda-\mu)+
\phi_1(\mu)-\phi_1(\lambda)\right) \right. \\
&+&\left.\exp\left(\alpha+\phi_3(\lambda)+\phi_4(\mu)\right)
\left(t(\mu, \lambda)+t(\lambda, \mu)\exp\left(-i x(\lambda-\mu)
+\phi_2(\lambda)-\phi_2(\mu)\right)\right)
\right\}.
\label{def:kernel-V}\nonumber
\end{eqnarray}
where
\begin{eqnarray}
t(\lambda,\mu)=\frac{(i c)^2}{(\lambda-\mu)(\lambda-\mu+i c)}.
\end{eqnarray}
We call the second-factor $\vartheta(\lambda)$ the Fermi weight :
\begin{eqnarray}
\vartheta(\lambda)=
\frac{1}{1+e^{\frac{\varepsilon(\lambda)}{T}}}
=\frac{\rho_p(\lambda)}{\rho_t(\lambda)}.
\end{eqnarray}
Because the dual fields $\phi_j(\lambda)$ commute with each other,
we can define the quantity
$\det\left(1+\frac{1}{2\pi}\hat{V}_T\right)$.
The integral operator $\hat{K}_T$ is defined by
\begin{eqnarray}
\left(\hat{K}_Tf\right)(\lambda)=
\int_{-\infty}^{\infty} K_T(\lambda, \mu)f(\mu)d\mu.
\end{eqnarray}
The integral kernel $K_T(\lambda,\mu)$ is defined
by $K_T(\lambda,\mu)=K(\lambda,\mu)\vartheta(\mu)$.
$K(\lambda,\mu)$ is defined in
(\ref{def:kernel-K}).
Now we state the results which we will use in the following sections.

\begin{thm}\cite{I.I.K.S.}~~~
In terms of the dual fields $\phi_j(\lambda)~
(j=1,\cdots,4),$ we can express the expectation value
$\langle \exp\left(\alpha Q(x)\right) \rangle_T$
by the Fredholm determinant :
\begin{eqnarray}
\langle \exp\left(\alpha Q(x)\right) \rangle_T
=\frac{\langle 0 \vert
\det \left(1+\frac{1}{2\pi}\hat{V}_T\right)\vert 0 \rangle}
{\det\left(1-\frac{1}{2 \pi}\hat{K}_T\right)}.
\label{eqn:expec-gen}
\end{eqnarray}
Here the symbol $\det\left(1+\frac{1}{2\pi}\hat{V}_T\right)$
represents the Fredholm determinant corresponding to
the following Fredholm integral equation of the second kind:
\begin{eqnarray}
\left(\left(1+\frac{1}{2\pi}\hat{V}_T\right)f\right)(\lambda)=g(\lambda),
~~\mbox{for}~\lambda \in (-\infty, \infty).
\end{eqnarray}
The denominator $\det\left(1-\frac{1}{2 \pi}\hat{K}_T\right)$
represents the Fredholm determinant corresponding to the
following Fredholm integral equation of the second kind :
\begin{eqnarray}
\left(\left(1-\frac{1}{2 \pi}\hat{K}_T\right)f\right)(\lambda)
=g(\lambda),~~\mbox{for}~\lambda \in (-\infty,\infty).
\end{eqnarray}
\end{thm}

\end{section}
\begin{section}{The small mass limit of the Bose particle}

In this section we will show that at the small mass limit :
$m \to 0,~g \to \infty,~\mbox{such that}~c=2mg~\mbox{is fixed}$,
a simplification occurs.
As explained in the Introduction, the scattering matrix
depends on the product $c=2mg$, not just on $g$.
Therefore the limit of small mass is not a free-fermion limit.
We want to emphasize this point.
In the sequel we consider the limit of small mass.
First we evaluate the solution of the Yang-Yang equation.
\begin{eqnarray}
\varepsilon(\lambda)&=&\frac{\lambda^2}{2m}
-h-\frac{T}{2\pi}\int_{-\infty}^{\infty}K(\lambda,\mu)\ln
\left(1+e^{-\frac{\varepsilon(\mu)}{T}}\right)d \mu.
\label{eqn:Y.Y.}
\end{eqnarray}
This is done following \cite{Y.Y.}.
\begin{lem}
~~~~~In the small mass limit of the Bose particle,
a solution of the Yang-Yang equation (\ref{eqn:Y.Y.}) is evaluated as :
\begin{eqnarray}
\varepsilon(\lambda)=\frac{\lambda^2}{2m}-h+O(\sqrt{m}).
\end{eqnarray}
\label{lem:eval}
\end{lem}
{\sl Proof}~~~~
In \cite{Y.Y.} C.N. Yang and C.P. Yang derived the following inequalities.
\begin{eqnarray}
\frac{\lambda^2}{2m}-h \geq \varepsilon(\lambda)
\geq \frac{\lambda^2}{2m}+x_0,
\end{eqnarray}
where $x_0$ is defined by the integral equation :
\begin{eqnarray}
x_0=-h-\frac{T}{2 \pi}\int_{-\infty}^{\infty}
K(0,\mu)\ln\left(1+e^{-\frac{1}{T}\left(\frac{\mu^2}{2m}+x_0\right)}\right)
du.
\end{eqnarray}
The existence of $x_0$ is proved in \cite{Y.Y.}.
Let us change the integration variable to $\nu=\frac{\mu}{\sqrt{2m}}$.
In the limit of small mass, $\sqrt{cg}$ tends to $\infty$.
Therefore we obtain : 
\begin{eqnarray}
x_0&=&-h-\frac{T}{\pi}\int_{-\infty}^{\infty}
\frac{\sqrt{cg}}{(\sqrt{cg})^2+\nu^2}
\ln \left(1+e^{-\frac{1}{T}\left(\nu^2+x_0\right)}\right)
d\nu\\
&=&-h+x_0-\frac{T}{\pi}\int_{-\infty}^{\infty}
\frac{\sqrt{cg}}{(\sqrt{cg})^2+\nu^2}
\ln \left(e^{\frac{1}{T}x_0}+e^{-\frac{1}{T}\nu^2}\right)
d\nu \label{eval2}\\
&=&-h-\frac{T}{\pi}\frac{1}{\sqrt{cg}}
\int_{-\infty}^{\infty}
\int_{-\infty}^{\infty}
\ln \left(1+e^{-\frac{1}{T}\left(\nu^2+x_0\right)}\right)
d\nu+O(m).\label{eval3}
\end{eqnarray}
When we assume $\vert x_0 \vert \to \infty$, this contradicts to
(\ref{eval2}).
Therefore we can assume that $\vert x_0 \vert$ is bounded.
Therefore, from the equation (\ref{eval3}), we can deduce 
$x_0=-h+O(\sqrt{m})$.

\hfill $\Box$

>From lemma \ref{lem:eval},
we can evaluate the Fermi weight $\vartheta(\lambda)$.
The Fermi weight $\vartheta(\lambda)$ 
has a very sharp maximum at $\lambda=0$,
from which it decreases to $0$ very fast.
Therefore a simplification occurs.
First we consider the dual fields.
In the sequel
we consider the case
that the spectral parameters are restricted to
$\lambda,\mu \approx O(\sqrt{m})$.
We observe the simplification of the commutation relations :
\begin{eqnarray}
[p_j(\lambda), q_k(\mu)]=
\left( \begin{array}{cccc}
0 & 0 & -1 & -1 \\
0 & 0 &  1 &  1 \\
1 &-1 &  0 &  0 \\
1 &-1 &  0 &  0
\end{array}\right)_{j,k}
\frac{i}{c}(\mu-\lambda)
+O(m).
\label{eqn:com-relation2}
\end{eqnarray}
Therefore we can identify pairs of fields :
\begin{eqnarray}
p_1(\lambda)=-p_2(\lambda),&& p_3(\lambda)=p_4(\lambda),\\
q_1(\lambda)=-q_2(\lambda),&& q_3(\lambda)=q_4(\lambda),\\
\phi_1(\lambda)=-\phi_2(\lambda),&&
\phi_3(\lambda)=\phi_4(\lambda).
\end{eqnarray}
Furthermore, because the first term of the commutation relation
(\ref{eqn:com-relation2}) is a linear function of the spectral parameters,
we can choose a representation of fields such that
$\phi_j(\lambda)$ are linear functions of the spectral parameter
$\lambda$ :
\begin{eqnarray}
\phi_j(\lambda)=\phi_j(0)+\phi_j'(0)\lambda,~~
\phi_j(0)=p_j(0)+q_j(0),~
\phi_j'(0)=p_j'(0)+q_j'(0),~~(j=1,3).
\label{eqn:identification}
\end{eqnarray}
Here the commutation relations are :
\begin{eqnarray}
[p_j(0),q_k(0)]=0=[p_j'(0),q_k'(0)],~~(j,k=1,3),
\end{eqnarray}
\begin{eqnarray}
[p_1'(0),q_3(0)]= \frac{i}{c} =-[p_3'(0),q_1(0)],~~
[p_3(0),q_1'(0)]= \frac{i}{c} =-[p_1(0),q_3'(0)].
\end{eqnarray}
The actions on the auxiliary vacuum are :
\begin{eqnarray}
p_1(0)\vert 0 \rangle=
p_3(0)\vert 0 \rangle=
p_1'(0)\vert 0 \rangle=
p_3'(0)\vert 0 \rangle=0,\\
\langle 0 \vert q_1(0)=
\langle 0 \vert q_3(0)=
\langle 0 \vert q_1'(0)=
\langle 0 \vert q_3'(0)=0
\end{eqnarray}
Furthermore we arrive at the following formula.
\begin{thm}
~~~~In the small mass limit of the Bose particle, 
the expectation value of the Fredholm determinant 
simplifies as follows:
\begin{eqnarray}
\langle 0 \vert \det
\left(1+\frac{1}{2\pi}\hat{V}_T\right) \vert 0\rangle
\longmapsto \langle 0 \vert \det
\left(1+\hat{V}_{0,T}\right) \vert 0 \rangle
.
\end{eqnarray}
Here the symbol $\hat{V}_{0,T}$ is the integral operator defined by
\begin{eqnarray}
\left(\hat{V}_{0,T}f\right)(\lambda)
=\int_{-\infty}^{\infty} V_{0,T}(\lambda,\mu)f(\mu)d\mu,
\end{eqnarray}
where the integral kernel is defined by product
\begin{eqnarray}
V_{0,T}(\lambda,\mu)=\left(\frac{e^{\hat{\alpha}}-1}{\pi}\right)
\frac{\sin\frac{\hat{x}}{2}(\lambda-\mu)}{\lambda-\mu}
\vartheta_0\left(\frac{\mu}{\sqrt{2mT}},\frac{h}{T}\right),
\end{eqnarray}
where 
\begin{eqnarray}
\vartheta_0(\mu,\beta)=\frac{1}{1+e^{\mu^2-\beta}}.
\label{eqn:mFermi}
\end{eqnarray}
Here we used the abbreviations:
\begin{eqnarray}
&&\hat{\alpha}=\alpha+\hat{\alpha}_p+\hat{\alpha}_q,~
\hat{x}=x+\hat{x}_p+\hat{x}_q,\\
&&\hat{\alpha}_p=2p_3(0),~\hat{\alpha}_q=2q_3(0),~
\hat{x}_p=-i p_1'(0),~\hat{x}_q=-i q_1'(0).
\end{eqnarray}
The commutation relations and the actions on
the auxiliary vacuum become:
\begin{eqnarray}
[\hat{x}_p,\hat{\alpha}_q]=\frac{2}{c},&&
[\hat{\alpha}_p,\hat{x}_q]=\frac{2}{c},\\
\hat{x}_p \vert 0 \rangle =0
=\hat{\alpha}_p \vert 0 \rangle,&&
\langle 0 \vert \hat{x}_q=0
=\langle 0 \vert \hat{\alpha}_q.
\end{eqnarray}
The dual fields $\hat{\alpha}$ and $\hat{x}$ commute with each other,
$[\hat{\alpha},\hat{x}]=0$.
\end{thm}

{\sl Proof.}
~~~~From lemma \ref{lem:eval}, the Fermi weight $\vartheta(\lambda)$
has a very sharp maximum at $\lambda=0$ and decrease to $0$
very fast. When we consider
the integral operator $\hat{V}_T$, we can restrict 
our consideration to the case of the spectral parameters
$\lambda, \mu \approx O(\sqrt{m})$.
Therefore we can use the above dual fields simplification.
We can identify four dual fields to two dual fields, which
are linear in the spectral parameters
$\lambda,\mu$.
Furthermore, since the relations $[p_3'(0),\phi_1(\lambda)-\phi_1(\mu)]=0,
~[q_3'(0),\phi_1(\lambda)-\phi_1(\mu)]=0$ and
$\langle 0 \vert q_3'(0)=0,~
p_3'(0)\vert 0 \rangle=0$ hold,
we can drop $p_3'(0),~q_3'(0)$ in
the expectation value $\langle 0 \vert \det
\left(1+\frac{1}{2\pi}\hat{V}_T\right) \vert 0 \rangle$.
Next we perform a similarity transformation
$\exp\left(\frac{i}{2}\lambda(x-i \phi_1'(0))\right)$
which leaves
the Fredholm determinant invariant.
Finally we substitute the Fermi weight
$\vartheta(\mu)$
by the modified
Fermi weight
$\vartheta_0\left(\frac{\mu}{\sqrt{2mT}},\frac{h}{T}\right)$.
We get the desired formula.
\hfill $\Box$

The denominator of
the expectation value
(\ref{eqn:expec-gen}) becomes the following :
\begin{eqnarray}
\det \left(1-\frac{1}{2\pi}\hat{K}_T \right)
=
1-\frac{\sqrt{2T}}{\pi c}
~d\left(\frac{h}{T}\right)\sqrt{m}+O(m),
\end{eqnarray}
where we used
\begin{eqnarray}
d(\beta)=\int_{-\infty}^{\infty}\vartheta_0(\mu,\beta)d\mu.
\label{def:d}
\end{eqnarray}
The density $D$ can be written as :
\begin{eqnarray}
D=\frac{N}{L}=\int_{-\infty}^{\infty}
\rho_p(\mu) d\mu=\frac{1}{2\pi}\int_{-\infty}^{\infty}
\frac{1}{1+e^{\frac{1}{T}\left(\frac{\mu^2}{2m}-h\right)}}d\mu
+O(m).
\end{eqnarray}
Therefore we can write
\begin{eqnarray}
\det \left(1-\frac{1}{2\pi}\hat{K}_T \right)
=
1-\frac{2}{c}D+O(m).
\end{eqnarray}
Therefore we arrive at the simplified formula for correlation functions.
\begin{cor}
~~~In the small mass limit of the Bose particle,
the temperature correlation function simplifies as follows :
\begin{eqnarray}
\langle \exp \left(\alpha Q(x)\right) \rangle_T
\longmapsto \langle 0 \vert \det
\left(1+\hat{V}_{0,T}\right) \vert 0\rangle
\left(1+\frac{2}{c}D\right).
\end{eqnarray}
Here $D=\frac{N}{L}$ is the density of the thermodynamic limit.
\label{cor:simp}
\end{cor}

\end{section}
\begin{section}{Maxwell-Bloch differential equation}

In this section we consider the differential equation
for the temperature correlation function in the small mass limit
of the Bose particle.
In the small mass limit, the Fredholm determinant
$\det\left(1+\hat{V}_{0,T}\right)$ is a $\tau$-function
of the Maxwell-Bloch equation, taking values in a commutative
subalgebra of the quantum operator algebra.
It is easily seen that after introducing
new variables, the auxiliary field $\hat{y}$ and the
scaled chemical potential
$\beta$,
\begin{eqnarray}
\hat{y}=y+\hat{y}_p+\hat{y}_q,~
y=\sqrt{\frac{mT}{2}}x,~
\hat{y}_p=\sqrt{\frac{mT}{2}}\hat{x}_p,~
\hat{y}_q=\sqrt{\frac{mT}{2}}\hat{x}_q,~
~~\beta=\frac{h}{T},
\end{eqnarray}
the Fredholm determinant
$\det\left(1+\hat{V}_{0,T}\right)$
can be rewritten, after the corresponding change 
$\lambda \to \frac{\lambda}{\sqrt{2mT}}$ of the spectral parameter, as
\begin{eqnarray}
\det\left(1+\hat{V}_{0,T}\right)=
\left.\det\left(1-\hat{\gamma} \hat{W}\right)
\right|_{\hat{\gamma}=\frac{1-\exp\left(\hat{\alpha}\right)}{\pi}}.
\end{eqnarray}
We want to emphasize that $\hat{y}$ is an operator in the auxiliary space.
The integral operator $\hat{W}$ is defined by
\begin{eqnarray}
\left(\hat{W}f\right)(\lambda)
=\int_{-\infty}^{\infty}W(\lambda,\mu)f(\mu)d\mu,
\end{eqnarray}
where the integral kernel
$W(\lambda,\mu)$ is given by
\begin{eqnarray}
W(\lambda,\mu)=\frac{\sin \hat{y}(\lambda-\mu)}{\lambda-\mu}
\vartheta_0(\mu,\beta).
\end{eqnarray}
The algebraic structure of the Fredholm determinant
$\left.\det\left(1-\hat{\gamma} \hat{W}\right)
\right|_{\hat{\gamma}=\frac{1-\exp(\hat{\alpha})}{\pi}}$
has been investigated in the context of
correlation functions for the impenetrable Bose gas \cite{K.B.I.}.
It is convenient to introduce the function $\sigma$ defined by
\begin{eqnarray}
\sigma \left(\hat{y},\beta,\hat{\alpha}\right)
=\left.\ln \det\left(1-\hat{\gamma} \hat{W} \right)
\right|_{\hat{\gamma}=\frac{1-\exp\left(\hat{\alpha}\right)}{\pi}}.
\end{eqnarray}
The operator $\sigma$ satisfies the
Maxwell-Bloch equation in the case that $\hat{y}$ and $\hat{\alpha}$
are real numbers \cite{K.B.I.}.
 In our case,
$\hat{y}$ and $\hat{\alpha}$ are quantum operator, but due to the fact that
they commute with each other, we can follow the derivation in \cite{K.B.I.}.
Therefore we arrive at the following results.
In the sequel we use the following operator-derivation notation :
\begin{eqnarray}
\frac{\partial}{\partial \hat{y}}F(\hat{y})
:=\left.\frac{\partial}{\partial z}F (z)\right|_{z=\hat{y}},
\end{eqnarray}
where $F=F(z)$ is a function of $z$.

\begin{prop}~~~
The operator $\sigma (\hat{y},\beta,\hat{\alpha})
=\left.\ln \det \left(1-\hat{\gamma}\hat{W}\right)
\right|_{\hat{\gamma}=\frac{1-\exp\left(\hat{\alpha}\right)}{\pi}}$
obeys the following nonlinear partial differential equation :
\begin{eqnarray}
\left(\frac{\partial}{\partial \beta}
\frac{\partial^2}{{\partial \hat{y}}^2} \sigma \right)^2
=-4 \left( \frac{\partial^2}{{\partial \hat{y}}^2} \sigma \right)
\left( 2\hat{y} \frac{\partial}{\partial \beta}
\frac{\partial}{\partial \hat{y}}\sigma+
\left(\frac{\partial}{\partial \beta} 
\frac{\partial}{\partial \hat{y}}\sigma \right)^2
-2\frac{\partial}{\partial \beta}\sigma \right),
\label{eqn:Max-Blo}
\end{eqnarray}
with the initial conditions :
\begin{eqnarray}
&&\sigma=-\left(\frac{1-e^{\hat{\alpha}}}{\pi} d(\beta)\right)~\hat{y}
-\left(\frac{1-e^{\hat{\alpha}}}{\pi} d(\beta)\right)^2\frac{\hat{y}^2}{2}
+O(\hat{y}^3),\\
&&\lim_{\beta \to -\infty}
\sigma(\hat{y},\beta,\hat{\alpha})=0,
\end{eqnarray}
where the scalar function $d(\beta)$ is defined in (\ref{def:d}).
\end{prop}

This initial data fixes the solution uniquely.
The nonlinear differential equation (\ref{eqn:Max-Blo})
is called the Maxwell-Bloch equation
\cite{B.G.Z.}.
Algebraically, it is known that at $T=0$ the operator
$\sigma$ depends only on product of variables 
$\hat{y}\sqrt{\beta}$ \cite{I.I.K.S.}. We set 
$\tau =\hat{y} \sqrt{\beta}
=\sqrt{\frac{mh}{2}}\hat{x}$.
Equation (\ref{eqn:Max-Blo}) is rewritten at $T=0$ 
for the operator
\begin{eqnarray}
\sigma_0(\tau)=\left.\tau \frac{d}{d \tau}
\ln \det\left(1-\hat{\gamma} \hat{W}\right)
\right|_{\hat{\gamma}=\frac{1-\exp\left(\hat{\alpha}\right)}{\pi}},
\end{eqnarray}
as
\begin{eqnarray}
\left(\tau \frac{d^2 \sigma_0}{d \tau^2}\right)^2
=-4\left(\tau \frac{d \sigma_0}{d \tau}-\sigma_0\right)
\left(4\tau \frac{d \sigma_0}{d \tau}
+\left(\frac{d \sigma_0}{d \tau}\right)^2
-4\sigma_0\right).\label{ord-Pain}
\end{eqnarray}
This ordinary differential equation is the
fifth Painlev\'e equation
in \cite{J.M.M.S.}.
Actually, rewriting (\ref{ord-Pain})
in terms of the function $y_0(\tau)$ defined by
\begin{eqnarray}
\sigma_0(\tau)=-4 i \tau u_0(\tau)
+\frac{u_0(\tau)^2}{y_0(\tau)}(y_0(\tau)-1)^2,~~
u_0(\tau)=\frac{4 i \tau y_0(\tau)-\tau \frac{dy_0(\tau)}{d \tau}}
{2(y_0(\tau)-1)^2},
\end{eqnarray}
we can get the familiar formula of the fifth Painlev\'e
differential equation for the function 
$w(\tau)=y_0\left(\frac{\tau}{2}\right)$ :
\begin{eqnarray}
\frac{d^2 w}{d\tau^2}=
\left(\frac{dw}{d\tau}\right)^2
\frac{3w-1}{2w(w-1)}+\frac{2w(w+1)}{w-1}+
\frac{2 i w}{\tau}-\frac{1}{\tau}\frac{d w}{d \tau}.
\end{eqnarray}
Next we derive the asymptotics of
$\sigma(\hat{y},\beta,\hat{\alpha})=
\left.\ln \det \left(1-\hat{\gamma} \hat{W} \right)
\right|_{\hat{\gamma}=\frac{1-\exp\left(\hat{\alpha}\right)}{\pi}}$.
By means of the Riemann-Hilbert method,
asymptotics of $\sigma$ are derived for the case when
$\hat{y}$ and $\hat{\alpha}$ are real number \cite{K.B.I.}.
The idea of the Riemann-Hilbert method
is due to Professor A.R. Its.
In our case, $\hat{y}$ and $\hat{\alpha}$ are quantum operator,
but due to the fact that they commute, we can
follow the derivation in \cite{K.B.I.}.
We arrive at the following asymptotics.
\begin{prop}~~~~
The asymptotics of the operator
$\sigma(\hat{y},\beta,\hat{\alpha})$ for large $\hat{y}$ become as follows :
\begin{eqnarray}
\sigma(\hat{y},\beta,\hat{\alpha})&=&
-\hat{y}~C(\beta,\hat{\alpha})+\frac{1}{2}
\int_{-\infty}^{\beta}\left(\frac{\partial C(b,\hat{\alpha})}
{\partial b}\right)^2 db +\\
&-&\frac{1}{8}\frac{(e^{-\hat{\alpha}}-1)^2}
{r_1(\hat{\alpha})^4~\vert
a(\lambda_1(\hat{\alpha}),\hat{\alpha}) \vert^4}
e^{-4r_1(\hat{\alpha}) \sin 
\varphi_1(\hat{\alpha})~\hat{y}} \nonumber \\
&\times&\left(
\frac{1}{\sin^2 \varphi_1(\hat{\alpha})}+\cos
\left\{4\hat{y}r_1(\hat{\alpha}) \cos 
\varphi_1(\hat{\alpha})
-4\arg a(\lambda_1(\hat{\alpha}),\hat{\alpha})
-4 \varphi_1(\hat{\alpha})\right\}\right)\nonumber \\
&+&o\left(e^{-4r_1(\hat{\alpha}) 
\sin \varphi_1(\hat{\alpha})~\hat{y}}\right).\nonumber 
\end{eqnarray}
Here we set
\begin{eqnarray}
&&C(\beta,\alpha)=\frac{1}{\pi}\int_{-\infty}^{\infty}
\ln \left(\frac{1+e^{\mu^2-\beta}}
{e^{\alpha} +e^{\mu^2-\beta}}\right) d \mu,\label{def:a1}\\
&&\lambda_1(\alpha)=\sqrt{\alpha+\beta+\pi i},~
r_1(\alpha)=\vert \lambda_1(\alpha) \vert,~\varphi_1(\alpha)=\arg 
\lambda_1(\alpha),\label{def:a2}\\
&&a(\lambda,\alpha)=\exp
\left\{\frac{1}{2 \pi i}\int_{-\infty}^{\infty}\frac{d \mu}
{\mu-\lambda} \ln
\left(\frac{1+e^{\mu^2-\beta}}{e^{\alpha}+e^{\mu^2-\beta}}
\right)\right\}.\label{def:a3}
\end{eqnarray}
\label{cor:asym1}
\end{prop}
\end{section}

\begin{section}{Evaluation of the mean value}

In this section we evaluate the vacuum expectation value
of the operator $\left.\det \left(1-\hat{\gamma} \hat{W}\right)
\right|_{\hat{\gamma}=\frac{1-\exp\left(\hat{\alpha}\right)}{\pi}}$ for
$y=\sqrt{\frac{mT}{2}}x \to +\infty$.
>From Corollary \ref{cor:asym1}, we deduce
\begin{eqnarray}
&&\langle 0 \vert
\left.\det \left(1-\hat{\gamma} \hat{W}\right)
\right|_{\hat{\gamma}=\frac{1-\exp\left(\hat{\alpha}\right)}{\pi}}
\vert 0 \rangle \nonumber \\
&=&\langle 0 \vert A\left(\beta,\hat{\alpha}\right)
e^{-C(\beta,\hat{\alpha})\hat{y}}
+
B\left(\beta,\hat{\alpha}\right)
e^{-\{C(\beta,\hat{\alpha})+4r_1(\hat{\alpha})\sin
\varphi_1(\hat{\alpha})\}\hat{y}}
\nonumber \\
&&~~~~+
G\left(\beta,\hat{\alpha}\right)
e^{\{-C(\beta,\hat{\alpha})+4i\lambda_1(\hat{\alpha})\}\hat{y}}+
H\left(\beta,\hat{\alpha}\right)
e^{\{-C(\beta,\hat{\alpha})-4i\lambda_1^{*}(\hat{\alpha})\}\hat{y}}
\vert 0 \rangle +\cdots.
\label{eqn:eval}
\end{eqnarray}
Here we set
\begin{eqnarray}
A(\beta,\alpha)
&=&\exp\left\{\frac{1}{2}\int^{\beta}_{-\infty}
\left(\frac{\partial C(b,\alpha)}{\partial b}\right)^2d b\right\},
\label{def:A}\\
B\left(\beta,\alpha \right)
&=&-\frac{(e^{-\alpha}-1)^2}
{8 r_1(\alpha)^4
\sin^2\varphi_1(\alpha)}
\left(\frac{a(\lambda_1^{*}(\alpha),\alpha)}
{a(\lambda_1(\alpha),\alpha)}\right)^2
A(\beta,\alpha),
\label{def:B}\\
G(\beta,\alpha)&=&-\frac{(e^{-\alpha}-1)^2}
{16}
\frac{1}{\lambda_1(\alpha)^4 a(\lambda_1(\alpha),\alpha)^4}
A(\beta,\alpha),
\label{def:G}\\
H(\beta,\alpha)&=&-\frac{(e^{-\alpha}-1)^2}
{16}
\frac{a(\lambda_1^{*}(\alpha),\alpha)^4}{\lambda_1^{*}(\alpha)^4}
A(\beta,\alpha),
\label{def:H}\end{eqnarray}
where $C(\beta,\alpha),
\lambda_1(\alpha),r_1(\alpha),\varphi_1(\alpha)$ and
$a(\lambda,\alpha)$ are defined in
(\ref{def:a1}), (\ref{def:a2}) and (\ref{def:a3}).
$\lambda_1^{*}(\alpha)$
is the complex conjugation of
$\lambda_1(\alpha)$, i.e.
\begin{eqnarray}
\lambda_1^{*}(\alpha)=\sqrt{\alpha+\beta-\pi i}.
\end{eqnarray}
$H(\beta,\alpha)$ is the complex conjugation of $G(\beta,\alpha)$.

In this section we evaluate the right-hand of the above 
vacuum expectation value.
For the reader's convenience we summarize the commutation relations
of the quantum operators :
\begin{eqnarray}
&&\hat{y}=y+\hat{y}_p+\hat{y}_q,~~
y=\sqrt{\frac{mT}{2}}x,~\hat{y}_p=\sqrt{\frac{mT}{2}}\hat{x}_p,~
\hat{y}_q=\sqrt{\frac{mT}{2}}\hat{x}_q,\\
&&\hat{\alpha}=\alpha+\hat{\alpha}_p+\hat{\alpha}_q,~~~
\beta=\frac{h}{T},
\end{eqnarray}
\begin{eqnarray}
[\hat{y}_p,\hat{\alpha}_q]=\frac{\sqrt{2mT}}{c}=
[\hat{\alpha}_p,\hat{y}_q],~~~
\hat{y}_p \vert 0 \rangle =0=\hat{\alpha}_p \vert 0 \rangle,~~
\langle 0 \vert \hat{y}_q=0=\langle 0 \vert \hat{\alpha}_q.
\end{eqnarray}
The following Proposition
is the key to calculating the vacuum expectation value.
\begin{prop}~~~
The following asymptotic formula holds at large $y\to +\infty$ :
\begin{eqnarray}
\langle 0 \vert
e^{\hat{y} E(\hat{\alpha})}
F(\hat{\alpha})
\vert 0 \rangle=
F\left(\alpha+\frac{\sqrt{2mT}}{c}E(\alpha)\right)
e^{ y E(\alpha)}+\cdots.
\label{eqn:key}
\end{eqnarray}
Here $E(\alpha)$ and $F(\alpha)$ are meromorphic functions of $\alpha$.
\label{prop:key}
\end{prop}
{\sl Proof.}
~~~~In this proof we use the following abbreviations :
\begin{eqnarray}
\delta=\frac{\sqrt{2mT}}{c},~~~A_0+A_1\hat{\alpha}_q+A_2
\hat{\alpha_q}^2+\cdots=E(\alpha+\hat{\alpha}_q).
\end{eqnarray}
First we expand the exponential function and use the relations
$\langle 0 \vert \hat{y}_q=0,~\hat{\alpha}_p \vert 0 \rangle=0$
and $[\hat{\alpha}_p, \hat{\alpha}_q]=0$.
We obtain
\begin{eqnarray}
(e.v.):=\langle 0 \vert
F(\hat{\alpha})\exp \left\{\hat{y}E(\hat{\alpha})\right\}
\vert 0 \rangle =
\langle 0 \vert \sum_{n=0}^{\infty}
\frac{1}{n!}(y+\hat{y}_p)^n (E(\alpha+\hat{\alpha}_q))^n
F(\hat{\alpha})\vert 0 \rangle.
\end{eqnarray}
We expand 
\begin{eqnarray}
(E(\alpha+\hat{\alpha}_q))^n
=(A_0+A_1\hat{\alpha}_q+A_2
\hat{\alpha_q}^2+\cdots)^n,
\end{eqnarray}
and using the commutation relation :
\begin{eqnarray}
\langle 0 \vert [f(\hat{y}_p),\hat{\alpha}_q^k]
=\delta^k \langle 0 \vert f^{(k)}(\hat{y}_p),
\end{eqnarray}
we obtain
\begin{eqnarray}
(e.v.)&=&
\langle 0 \vert
\sum_{n=0}^{\infty}\frac{1}{n!}(y+\hat{y}_p)^n
\sum_{m_0+m_1+m_2+\cdots=n \atop{m_j \geq 0}}
\frac{n!}{m_0! m_1! m_2! \cdots }
A_0^{m_0}A_1^{m_1}A_2^{m_2}\cdots \nonumber \\
&\times& \hat{\alpha}_q^{m_1+2m_2+3m_3+\cdots}
F(\hat{\alpha})
\vert 0 \rangle \\
&=&
\langle 0 \vert
\sum_{n=0}^{\infty}
\sum_{m_0+m_1+m_2+\cdots=n \atop{m_j \geq 0}}
\frac{n!}{m_0! m_1! m_2! \cdots }~\delta^{m_1+2m_2+3m_3+\cdots}
(y+\hat{y}_p)^{n-(m_1+2m_2+3m_3+\cdots)}\nonumber \\
&\times& A_0^{m_0}A_1^{m_1}A_2^{m_2}\cdots
\frac{n(n-1)\cdots(n+1-(m_1+2m_2+3m_3+\cdots))}{n!}F(\hat{\alpha})
\vert 0 \rangle.
\end{eqnarray}
Using the relation 
\begin{eqnarray}
\frac{1}{2\pi i}\oint \frac{e^t}{t^{n-k+1}}dt
=\frac{n(n-1)\cdots(n-k+1)}{n!}
=\frac{1}{(n-k)!},
\end{eqnarray}
we can factor as follows :
\begin{eqnarray}
(e.v.)
&=&\langle 0 \vert \sum_{n=0}^{\infty}
\frac{1}{2 \pi i}\oint \frac{e^t}{t}
\left(\frac{A_0(y+\hat{y}_p)}{t}
+A_1 \delta +\frac{A_2 \delta^2 t}{y+\hat{y}_p}
+\frac{A_3 \delta^3 t^2}{(y+\hat{y}_p)^2}+\cdots\right)^n
F(\hat{\alpha})\vert 0 \rangle \\
&=&
\langle 0 \vert \sum_{n=0}^{\infty}
\frac{1}{2 \pi i}\oint \frac{e^t}{t}
\left(\frac{A_0(y+\hat{y}_p)}{t}\right)^n
F(\hat{\alpha})\vert 0 \rangle + \cdots ~~~{\rm for}~~~y \to +\infty.
\end{eqnarray}
Using the relations :
\begin{eqnarray}
\frac{1}{1-z}=1+z+z^2+z^3+\cdots,~~~f(z)=\frac{1}{2 \pi i}
\oint \frac{f(t)}{t-z} dt,
\end{eqnarray}
we get the following :
\begin{eqnarray}
(e.v.)=
\langle 0 \vert 
\frac{1}{2 \pi i}\oint \frac{e^t}{t-A_0(y+\hat{y}_p)}
F(\hat{\alpha})\vert 0 \rangle +\cdots
=
\langle 0 \vert
e^{A_0(y+\hat{y}_p)}
F(\hat{\alpha})\vert 0 \rangle +\cdots.
\end{eqnarray}
Using the relation
$e^A B e^{-A}=e^{ad(A)}(B)$,
we obtain
\begin{eqnarray}
e^{E(\alpha)\hat{y}_p}F(\hat{\alpha})
e^{-E(\alpha)\hat{y}_p}
=e^{E(\alpha)ad(\hat{y}_p)}F(\hat{\alpha})
=\left.e^{E(\alpha)\frac{\sqrt{2mT}}{c}
\frac{\partial}{\partial a}}
F(a)\right|_{a=\hat{\alpha}}
\end{eqnarray}
Therefore we can drop the quantum operators in
the expectation value.
\begin{eqnarray}
(e.v.)=e^{y E(\alpha)}
\langle 0 \vert \left.e^{E(\alpha)
\frac{\sqrt{2mT}}{c} \frac{\partial}{\partial a}}
F(a)\right|_{a=\hat{\alpha}}
e^{E(\alpha)\hat{y}_p}\vert 0 \rangle +\cdots
=\left.e^{E(\alpha)
\frac{\sqrt{2mT}}{c} \frac{\partial}{\partial a}}
F(a)\right|_{a=\alpha}+\cdots.
\label{def:D}
\end{eqnarray}
Here we have used the relations
$\langle 0 \vert \hat{\alpha}_q=0
=\hat{\alpha}_p \vert 0 \rangle,~~
\hat{y}_p \vert 0 \rangle=0$.

Because the exponential of derivation is a shift operator :
\begin{eqnarray}
e^{w \frac{\partial}{\partial z}}f(z)=f(z+w),
\end{eqnarray}
we arrive at (\ref{eqn:key}).

\hfill $\Box$

Now, we arrive at the following theorem.
\begin{thm}~~~~
The leading terms of the asymptotics of the expectation value
behave exponentially as follows.
\begin{eqnarray}
&&\langle 0 \vert
\left.\det \left(1-\hat{\gamma} \hat{W}\right)
\right|_{\hat{\gamma}=\frac{1-\exp\left(\hat{\alpha}\right)}{\pi}}
\vert 0 \rangle \\
&=&A\left(\beta,\alpha-\frac{\sqrt{2mT}}{c}C(\beta,\alpha)\right)
e^{-C(\beta,\alpha)y}\nonumber\\
&+&
B\left(\beta,\alpha-\frac{\sqrt{2mT}}{c}
\left\{C(\beta,\alpha)+4r_1(\alpha)
\sin
\varphi_1(\alpha)\right\}\right)
e^{-\{C(\beta,\alpha)+4r_1(\alpha)\sin
\varphi_1(\alpha)\}y}
\nonumber \\
&+&
G\left(\beta,\alpha+\frac{\sqrt{2mT}}{c}\left\{
-C(\beta,\alpha)+4i\lambda_1(\alpha)\right\}\right)
e^{\{-C(\beta,\alpha)+4i\lambda_1(\alpha)\}y}\nonumber \\
&+&
H\left(\beta,\alpha+\frac{\sqrt{2mT}}{c}
\{-C(\beta,\alpha)-4i\lambda_1^{*}(\alpha)\}\right)
e^{\{-C(\beta,\alpha)-4i\lambda_1^{*}(\alpha)\}y}+\cdots.
\nonumber
\end{eqnarray}
Here
$A(\beta,\alpha),B(\beta,\alpha),G(\beta,\alpha)$
and $H(\beta,\alpha)$ are defined in
(\ref{def:A}),(\ref{def:B}),(\ref{def:G}) and (\ref{def:H}).
\label{thm:asymp1}
\end{thm}
{\sl Proof}
~~~~Applying Proposition \ref{prop:key}
to (\ref{eqn:eval}), we arrive at the result.
\hfill $\Box$

When we consider $c=\infty$,
Theorem \ref{thm:asymp1} coincides with the asymptotics results 
for the impenetrable Bose gas
case \cite{K.B.I.}.

\begin{cor}~~~~
At the limit
$m \to 0,~g \to \infty,~x \to \infty$
such that $c=2mg$ fixed and $\sqrt{m}x \to \infty$,
 the leading terms of asymptotics of
the expectation value
become :
\begin{eqnarray}
\langle j(x)j(0) \rangle_T &\to&
D^2+\\
&+&\frac{mT}{2}\left(
B_0(\beta)+B_1(\beta)\sqrt{\frac{mT}{2}}x
+B_2(\beta)\left(\sqrt{\frac{mT}{2}}x\right)^2\right)
\exp\left\{-4r_1(0)\sin\varphi_1(0)
\sqrt{\frac{mT}{2}}x\right\}\nonumber \\
&+&\frac{mT}{2}\left(
G_0(\beta)+G_1(\beta)\sqrt{\frac{mT}{2}}x
+G_2(\beta)\left(\sqrt{\frac{mT}{2}}x\right)^2\right)
\exp\left\{4i\lambda_1(0)
\sqrt{\frac{mT}{2}}x\right\}\nonumber \\
&+&\frac{mT}{2}\left(
H_0(\beta)+H_1(\beta)\sqrt{\frac{mT}{2}}x
+H_2(\beta)\left(\sqrt{\frac{mT}{2}}x\right)^2\right)
\exp\left\{-4i\lambda_1^{*}(0)
\sqrt{\frac{mT}{2}}x\right\}+\cdots.\nonumber
\end{eqnarray}

Here $D=\frac{N}{L}$ is the density of the thermodynamics
and $\beta=\frac{h}{T}$.
Here $B_j(\beta),~G_j(\beta)$ and $H_j(\beta),~(j=0,1,2)$ are
functions of $\beta$. $H_j(\beta)~(j=0,1,2)$ is the complex
conjugation of $G_j(\beta)~(j=0,1,2)$, i.e.
$H_j(\beta)=G_j^{*}(\beta)$. 
Explicit formulas for $B_j(\beta),~G_j(\beta)$ and $H_j(\beta)~(j=0,1,2)$
are summarized in the Appendix.
\label{cor:asymp2}
\end{cor}
{\sl Proof.}
~~~~~From Corollary \ref{cor:simp},
Corollary \ref{cor:asymp2} and the relation
\begin{eqnarray}
\langle j(x)j(0) \rangle_T
=\frac{1}{2}
\frac{\partial^2}{\partial x^2}
\langle Q(x)^2 \rangle_T
=\frac{1}{2}\frac{\partial^2}{\partial x^2}
\left.\frac{\partial^2}{\partial \alpha^2}
\langle \exp \left(\alpha Q(x)\right)\rangle_T
\right|_{\alpha=0},
\end{eqnarray}
we can derive the result.
For example the constant $D^2$ is derived by :
\begin{eqnarray}
D^2=\left(1+\frac{2}{c}D\right)\frac{1}{2}
\frac{\partial^2}{\partial x^2}
\frac{\partial^2}{\partial \alpha^2}\left.
A\left(\beta,\alpha-\frac{\sqrt{2mT}}{c}C(\beta,\alpha)\right)
e^{-C(\beta,\alpha)y}\right|_{\alpha=0}+\cdots.
\end{eqnarray}
\hfill $\Box$

V. Korepin \cite{K} proposed a method of presenting correlation functions
in the form of special series.
This method is useful in the  calculation
of the long distance asymptotics.
N. Bogoliubov and V. Korepin \cite{B.K.} considered the asymptotics
of correlation functions for the penetrable Bose gas
by the special series method.
Corollary \ref{cor:asymp2} coincides with the result of \cite{B.K.}.
For the impenetrable Bose gas case $(c=\infty)$,
V. Korepin and N. Slavnov \cite{K.S.} calculated higher order corrections
and derived pre-exponential
polynomials by the special series method.
In this paper we derived pre-exponential polynomials for penetrable
Bose gas case $(0<c<+\infty)$ by using the determinant representation.

\end{section}

\section*{Appendix}

In this Appendix we summarize
the asymptotics of the density-density correlation function.
We use the notation given in Corollary \ref{cor:asymp2}.
In the sequel, we use the following abbreviations :
\begin{eqnarray}
\lambda_1=\lambda_1(0)=\sqrt{\beta +\pi i},~
r_1=\vert \lambda_1 \vert=\vert \lambda_1(0) \vert,~
\varphi_1=\arg \lambda_1=\arg \lambda_1(0).
\end{eqnarray}
First we summarized the coefficients of
$\exp\left\{-4r_1\sin \varphi_1 \sqrt{\frac{mT}{c}}x\right\}$ :

\begin{eqnarray}
\frac{mT}{2}\left(
B_0(\beta)+B_1(\beta)\sqrt{\frac{mT}{2}}x
+B_2(\beta)\left(\sqrt{\frac{mT}{2}}x\right)^2\right)
\exp\left\{-4r_1\sin\varphi_1
\sqrt{\frac{mT}{2}}x\right\}.
\end{eqnarray}
The functions $B_j(\beta)$ are given by
\begin{eqnarray}
B_j(\beta)=\tilde{B}_j(\beta)\left(1+\frac{2}{c}D\right),~~(j=0,1,2),
\end{eqnarray}
where $\tilde{B}_j(\beta)$ are given by

\begin{eqnarray}
\tilde{B}_2(\beta)&=& 
16r_1^2\sin^2 \varphi_1\left(\frac{d(\beta)}{\pi}+
\frac{2\sin \varphi_1}{r_1} \right)^2\times
B\left(\beta,-4\Delta r_1 \sin \varphi_1\right),\\
\tilde{B}_1(\beta)&=&
32 r_1^2 \sin^2 \varphi_1
\left(\frac{d(\beta)}{\pi}+\frac{2\sin \varphi_1}{r_1}\right)
\left(1+\Delta \left(\frac{d(\beta)}{\pi}+
\frac{2 \sin \varphi_1}{r_1}\right)\right)
\times
\left(\frac{\partial B}{\partial \alpha}\right)
\left(\beta,-4\Delta r_1 \sin \varphi_1\right)\nonumber \\
&-&4r_1\sin \varphi_1
\left\{\left(\frac{2d(\beta)}{\pi}\right)^2
+\frac{16\sin \varphi_1}{\pi r_1}d(\beta)
-\frac{4r_1 \sin \varphi_1}{\pi}\left(\frac{\partial d}{\partial \beta}\right)
(\beta)\right.\\
&+& \left.\frac{4\sin^2\varphi_1}{r_1^2}(5+\cos 2\varphi_1) \right\}
\times 
B(\beta,-4\Delta r_1 \sin \varphi_1)\nonumber \\
\tilde{B}_0(\beta)&=&
16 r_1^2 \sin^2 \varphi_1
\left\{1+\Delta \left(\frac{d(\beta)}{\pi}+\frac{2 \sin \varphi_1}{r_1}
\right)\right\}^2\times\left(\frac{\partial^2 B}{\partial \alpha^2}\right)
(\beta,-4\Delta r_1 \sin \varphi_1)\nonumber \\
&-&
4r_1\sin \varphi_1
\left[
\frac{4d(\beta)}{\pi}+\frac{8 \sin \varphi_1}{r_1}+
\Delta \left\{\left(\frac{2d(\beta)}{\pi}\right)^2
+\frac{16 \sin \varphi_1}{\pi r_1}d(\beta)\right. \right. \\
&-&\left. \left. \frac{4 r_1\sin \varphi_1}{\pi}
\left(\frac{\partial d}{\partial \beta} \right)(\beta)
+\frac{4 \sin^2 \varphi_1}{r_1^2}
(5+\cos 2\varphi_1)\right\}
\right]\times\left(\frac{\partial B}{\partial \alpha}\right)
(\beta,-4\Delta r_1 \sin \varphi_1)\nonumber \\
&+&2 
\left\{
\left(\frac{d(\beta)}{\pi}\right)^2
+\frac{4 \sin \varphi_1}{\pi r_1}d(\beta)
-\frac{4r_1 \sin \varphi_1}{\pi}
\left(\frac{\partial d}{\partial \beta}\right)(\beta)
+\frac{4 \sin^2 2 \varphi_1}{r_1^2}\right\}\times 
B\left(\beta,-4\Delta r_1 \sin \varphi_1\right).\nonumber 
\end{eqnarray}
Here we used the abbreviations :
\begin{eqnarray}
\Delta=\frac{\sqrt{2mT}}{c},~d(\beta)=\int_{-\infty}^{\infty}
\frac{1}{1+e^{\mu^2-\beta}}d \mu,~\beta=\frac{h}{T},
\end{eqnarray}
and the function $B(\beta,\alpha)$ as defined in (\ref{def:B}).

Next we summarize the coefficients of
$\exp\left\{4i\lambda_1 \sqrt{\frac{mT}{2}}x\right\}$ :
\begin{eqnarray}
\frac{mT}{2}\left(
G_0(\beta)+G_1(\beta)\sqrt{\frac{mT}{2}}x
+G_2(\beta)\left(\sqrt{\frac{mT}{2}}x\right)^2\right)
\exp\left\{4i \lambda_1\sqrt{\frac{mT}{2}}x\right\}
\end{eqnarray}
The functions $G_j(\beta)$ are given by

\begin{eqnarray}
G_j(\beta)=\tilde{G}_j(\beta)\left(1+\frac{2}{c}D\right),~~(j=0,1,2).
\end{eqnarray}
where $\tilde{G}_j(\beta)$ are given by
\begin{eqnarray}
\tilde{G}_2(\beta)&=&
-16 \lambda_1^2\left(\frac{d(\beta)}{\pi}+
\frac{2i}{\lambda_1} \right)^2\times
G\left(\beta,4\Delta i \lambda_1\right),\\
\tilde{G}_1(\beta)&=&
- 32 \lambda_1^2
\left(\frac{d(\beta)}{\pi}+\frac{2i}{\lambda_1}\right)
\left(1+\Delta \left(\frac{d(\beta)}{\pi}+
\frac{2i}{\lambda_1}\right)\right)
\times \left(\frac{\partial G}{\partial \alpha}\right)
\left(\beta,4\Delta i\lambda_1\right)\\
&+&4i \lambda_1
\left\{\left(\frac{2d(\beta)}{\pi}\right)^2
+\frac{16i}{\pi \lambda_1}d(\beta)
+\frac{4i\lambda_1}{\pi}\left(\frac{\partial d}{\partial \beta}\right)
(\beta)
-\frac{12}{\lambda_1^2}\right\}
\times G(\beta,4\Delta i\lambda_1 ),\nonumber \\
\tilde{G}_0(\beta)&=&
-16 \lambda_1^2
\left\{1+\Delta \left(\frac{d(\beta)}{\pi}+\frac{2i}{\lambda_1}
\right)\right\}^2\times\left(\frac{\partial^2 G}{\partial \alpha^2}\right)
(\beta,4\Delta i \lambda_1) \\
&+&
4 i \lambda_1
\left[
\frac{4d(\beta)}{\pi}+\frac{8 i}{\lambda_1}+
\Delta \left\{\left(\frac{2d(\beta)}{\pi}\right)^2
+\frac{16 i}{\pi \lambda_1}d(\beta)+ \frac{4 i \lambda_1}{\pi}
\left(\frac{\partial d}{\partial \beta} \right)(\beta)
-\frac{12}{\lambda_1^2}\right\}
\right]\times\left(\frac{\partial G}{\partial \alpha}\right)
(\beta,4\Delta i \lambda_1)\nonumber \\
&+&2
\left\{
\left(\frac{d(\beta)}{\pi}\right)^2
+\frac{4i}{\pi \lambda_1}d(\beta)
+\frac{4i \lambda_1}{\pi}
\left(\frac{\partial d}{\partial \beta}\right)(\beta)\right\}
\times G\left(\beta,4\Delta i \lambda_1 \right),\nonumber
\end{eqnarray}
Here the function $G(\beta,\alpha)$ is defined in (\ref{def:G}).
Next we summarize the coefficients of
$\exp\left\{-4i\lambda_1^{*} \sqrt{\frac{mT}{2}}x\right\} :$
\begin{eqnarray}
\frac{mT}{2}\left(
H_0(\beta)+H_1(\beta)\sqrt{\frac{mT}{2}}x
+H_2(\beta)\left(\sqrt{\frac{mT}{2}}x\right)^2\right)
\exp\left\{-4i \lambda_1^*
\sqrt{\frac{mT}{2}}x\right\}
\end{eqnarray}
where $\lambda_1^*=\sqrt{\beta-\pi i}$.
The functions $H_j(\beta)$ are given by
the complex conjugation of $G_j(\beta)$.
\begin{eqnarray}
H_j(\beta)=G_j^{*}(\beta),~~(j=0,1,2).
\end{eqnarray}

\section*{Acknowledgments}
We wish to thank Dr. J.-U.H. Petersen for correcting the English
of our paper.
This work is partly supported by the National Science Foundation
(NSF) under Grants No. PHY-9321165
and the Japan Society for the Promotion
of Science.


\begin{thebibliography}{99}
\bibitem{K.B.I.}Korepin V.E., Bogoliubov N.M. and
Izergin A.G. :
{\it Quantum Inverse Scattering Method and Correlation
Functions}, Cambridge Monographs
on Mathematical Physics 1993.
\bibitem{L.L.}Lieb E.H. and Liniger W. :
Exact analysis of an interacting Bose gas I.
The general solution and the ground state,
{\it Phys. Rev.} {\bf 130}, 1605-16 (1963)
\bibitem{L}Lieb E.H. :
Exact analysis of an interacting Bose gas II.
The excitation spectrum,
{\it Phys. Rev.} {\bf 130}, 1616-34 (1963)
\bibitem{Y.Y.}Yang C.N. and Yang C.P. :
Thermodynamics of a one dimensional systems of bosons
with repulsive delta-function interactions,
{\it J. Math. Phys.} {\bf 10}, 1152-1122 (1969)
\bibitem{B.G.Z.}Burtsev S.P., Gatitov I.R. and Zakharov V.E.:
Maxwell-Bloch system with pumping,
in {\it Plasma Theory and Nonlinear and Turbulent Processes in
Physics vol.2}, 897-905, World Scientific: Singapore 1988.
\bibitem{I.I.K.S.}Korepin V.E. :
Generating functional of correlation functions for the nonlinear
Schr\"odinger equation,
{\it Funct. Analiz. i jego Prilozh.} {\bf 23}, 15-23, (1989)~
(in Russian)
\bibitem{J.M.M.S.}Jimbo M., Miwa T., M\^ori Y.
and Sato M. :
Density matrix of an impenetrable Bose gas and the fifth
Painlev\'e transcendnt,
{\it Physica} {\bf 1D}, 80-158 (1980)
\bibitem{K}Korepin V.E. :
Correlation Functions of the One dimensional Bose Gas in
thr Repulsive Case,
{\it Commmun. Math. Phys.} {\bf 94}, 93-113 (1984)
\bibitem{B.K.}N.M. Bogoliubov and V.E. Korepin :
Correlation length of the one-dimensional Bose gas,
{\it Nucl. Phys.} {\bf B [FS14]}, 766-778 (1985)
\bibitem{K.S.}Korepin V.E. and Slavnov N.A. :
Correlation function of currents in
a one-dimensional Bose gas,
{\it Theor. Math. Phys.} v.{\bf 68}, 955-960 (1986)
\bibitem{F.J.}Frenkel I.B. and Jing N. :
Vertex representations of quantum affine algebras,
{\it Proc. Natl. Acad. Sci. USA} {\bf 85}, 9373-9377 (1988)
\bibitem{E.F.I.K.2}Essler F.H., Frahm H., Its A.R. and
Korepin V.E. :
Painlev\'e Transcendent Describes Quantum Correlation
Function of the XXZ Antiferromagnet away from the free-fermion
point, (1996), Yukawa-preprint YITP-96-13, solv-int/9604005.
\end{thebibliography}
\end{document}